\documentclass{elsart}

 \usepackage{epsfig}

\usepackage{amssymb}

\begin{document}

\begin{frontmatter}



\title{Mechanical Stretching of Proteins: Calmodulin and Titin}


\author[au1]{Marek Cieplak} 

\address[au1]{Institute of Physics, Polish Academy of Sciences,
Al. Lotnik\'ow 32/46, 02-668 Warsaw, Poland}

\begin{abstract}
Mechanical unfolding of several domains of calmodulin and titin
is studied using a Go-like model with a realistic
contact map and Lennard-Jones contact interactions. 
It is shown that this simple model captures the experimentally
observed difference between the two proteins: titin is a spring that
is tough and strong whereas calmodulin acts like a weak spring with 
featureless force-displacement curves. The difference is related
to the dominance of the $\alpha$ secondary structures in the
native structure of calmodulin. The tandem arrangements of calmodulin
unwind simultaneously in each domain whereas the domains in titin
unravel in a serial fashion. 
The sequences of contact events during unraveling are
correlated with the contact order, i.e. with the separation
between contact making amino acids along the backbone in the native state.
Temperature is found to affect stretching in a profound way.

\end{abstract}

\begin{keyword}
Protein stretching \sep molecular dynamics \sep Go model \sep calmodulin

\PACS 87.10.+e \sep 87.15.-v
\end{keyword}
\end{frontmatter}

\section{Introduction}
\label{1}

Recent developments in technology have enabled studies of large
biomolecules through mechanical manipulation. The most common
techniques od such a manipulation involve atomic force microscopy
and optical tweezers. These techniques target the 
hydrophobic and hydrogen bond interactions that are an order of
magnitude weaker than those corresponding to the covalent bonds.
The simplest protocol used in the studies 
is to anchor one segment of a molecule to
a substrate and pull by another segment at a constant speed, $v_p$.
By monitoring the force of resistance to the pull, $F$, and plotting
it versus the tip displacement, $d$, one obtains an elastic
characterization of a molecule that requires theoretical
interpretation in terms of the sequence of the rupturing events.
The resulting $F-d$ plots have
a fine structure consisting of peaks, minima, and plateaus that depend
on the molecule and conditions of its environment. The first
systems that were studied through stretching
were the streptavidin-biotin complex \cite{Florin,Grubmueller}, 
DNA \cite{Smith,Cluzel,Heslot}, and the 
multi-domained titin \cite{Gaub,Simmons,Kellermayer} 
that is found in a class of sarcomers
in muscles.

Pulling apart two strands of the DNA involves breaking one hydrogen
bond at a time and this makes the force ondulate around the value of 13 pN
and with the amplitude not exceeding 1 pN \cite{Heslot}. The
typicall pulling speed in these experiments is 40 nm/s and a five-fold
increase in $v_p$ affects the $F-d$ pattern only very weakly.
Separating biotin from streptavidin involves stretching many bonds
simultaneously and, within the pulling distance of about 10 \AA, the peak
force is close to 300 pN \cite{Grubmueller}. Stretching of titin results
in a sawtoothlike pattern \cite{Gaub,Clarke,Marszalek0,Marszalek}
where each tooth is 
attributed to unwinding
of a single domain and has a peak value of order 200 pN.
Titin has been found to act like a non-Hookean spring
which is strong, tough and nearly reversible \cite{Clarke}.
Its properties have inspired biomimetic design of polymers \cite{Guan}.
The calcium binding C2A protein, on the other hand, has been
found to be much weaker: the peak force is only of order 60 pN and
its sawtoothlike pattern has more structure
within each tooth 
\cite{Progress}. For poly-calmodulin, another calcium binding protein,
there are no significant force peaks since no cluster of hydrogen bonds
undergoes breaking \cite{Progress} until the molecule is fully stretched.
The patterns obtained for modular proteins generally depend on 
whether the modules are connected end-to-end or away from the terminals, 
like in the case of ubiquitin \cite{Carrion-Vazquez} or lyzozyme 
\cite{Yang,entropic}.

The $F-d$ patterns appear to act like finger-prints of biomolecules but
they also depend on the temperature, $T$, as evidenced
experimentally \cite{Janoviak} and theoretically \cite{thermtit,entropic}.
In this paper, we demonstrate that simple geometry based
theoretical models can capture substantial differences 
in the $F-d$ curves that exist between proteins and show that
these differences diminish on increasing the temperature and disappear
in the entropic limit. Our presentation is focused on two proteins:
the I27 domain of titin and calmodulin corresponding to the
Protein Data Bank \cite{PDB} codes 1tit and 1cfc respectively.
The former has the architecture
of the $\beta$ sandwich \cite{Pastore} with no $\alpha$ helices
and with the $\beta$ content of 32.65 \% whereas calmodulin is mostly
an $\alpha$ protein that shown in figure 1:
the helical content is 54.05 \% and the $\beta$
content is 8.11 \%.

There are obvious advantages to all-atom modelling compared to
simplified coarse grained models: it offers a more realistic
description and its nature is more fundamental. 
All atom modelling of 1tit \cite{Lu} leads to the identification
of the hydrogen bonds linking the so called A' and G strands
as being responsible for the maximum force in this case.
There are also equally clear drawbacks that are
related to the necessity of dealing only with very short time scales,
typically of order nano-seconds. This results in considering
the pulling speeds which are 6--7 orders of magnitude too rapid 
and which may be responsible for an order of magnitude too
big peak forces calculated for titin \cite{Lu} (another reason
for the discrepancy with the experiment could be the surface tension
effects due to the droplet of water that surrounds the protein).
The models presented here allow for studies that a) involve
more realistic $v_p$, b) incorporate tandem connection of
several domains, c) enable comparison to the kinetics of folding,
d) deal with variations of parameters, such as the $T$, and d)
easily compare various proteins.

\section{Model}

Protein folding is thought to be governed
by the geometry of the protein \cite{geom,tubes,tubes1,Du}
and especially by the geometry
of its native state \cite{Unger,Plaxco,Plaxco1}.
One way to incorporate geometry into the model is to follow
the prescription of Go \cite{Goabe,Stakada}: construct a Hamiltonian
that incorporates the chain-like connectivity and which has a 
ground state that agrees with the experimentally
determined native conformation. 
Our realisation of this prescription within a coarse grained model
that is studied through the techniques of molecular dynamics
is outlined in references \cite{Hoang,Hoang1,biophysical}. 

Briefly, the amino acids
are represented by point particles of mass $m$ located at the
positions of the C$^{\alpha}$ atoms. They are tethered by a strong
harmonic potential with a minimum at 3.8 \AA. 
The interactions between the
amino acids are grouped into contacts of the native and non-native kinds.
The distinction is based on taking the atomic representation
of the amino acids in the native state and then checking for 
their posible overlaps. The occurrence of an overlap is determined
assuming that the atoms take a spherical space corresponding to
the van der Waals radii of the atoms, enlarged
by the factor of 1.24 \cite{Tsai,prion} to account for the
soft part of the interaction potential.
The amino acids ($i$ and $j$) that are found to overlap in this
sense are considered to be forming contacts. These pairs are
endowed with the Lennard-Jones  potential, 
\begin{equation}
V_{ij} =
4\epsilon \left[ \left( \frac{\sigma_{ij}}{r_{ij}}
\right)^{12}-\left(\frac{\sigma_{ij}}{r_{ij}}\right)^6\right] \;,
\end{equation}
such that
its minimum agrees with the experimental value of the distance
between the C$^{\alpha}$ atoms in the native state.
This condition selects a pair by pair value of the length
parameter $\sigma _{ij}$ whereas the energy parameter $\epsilon$
is kept uniform. $\epsilon$ could be made specific if understanding
regarding the values was reached. It corresponds to many effective
non-covalent interactions, such as hydrophobicity and hydrogen bonds,
so it should range between 800 and 2300 K. It appears \cite{thermtit}
that $\tilde{T} = k_BT/\epsilon$ of about 0.3, 
where $k_B$ is the Boltzmann constant, qualitatively reproduces
the room temperature elastic behavior of titin.

The properties of the native contacts in the two
proteins studied here are illustrated in Figure 2.
There are 209 contacts (89 amino acids) in 1tit and 426 contacts
(148 amino acids) in 1cfc. For 1cfc, 71 \% of the contacts are
local -- their sequence distance does not exceed 4. This reflects
the high $\alpha$-content. In contrast, only 30 \% of the contacts in 
1tit are local. The contact map for 1tit, also shown in Figure 2, is
organised in stripes corresponding to interactions between 
distinct $\beta$-strands. Such patterns are not present in the case
of 1cfc suggesting a qualitatively different network of the couplings.

The thermal fluctuations away form the native state are mimicked
in the molecular dynamics simulation by introducing the Langevin 
noise with the damping constant $\gamma$
of 2 $m/\tau$, where $\tau$ is $\sqrt{m \sigma ^2 /\epsilon}$.
This corresponds to the situation in which the inertial effects are
negligible  \cite{biophysical}
but a more realistic account of the water environment
requires $\gamma$ to be about 25 times larger \cite{Veitshans}. Thus
the times scales obtained for $\gamma$=2$m/\tau$ need to be multiplied
by 25 since a linear dependence
on $\gamma$ has been found \cite{Hoang,Hoang1}.

Stretching is implemented by attaching both ends of the protein to harmonic
springs of spring constant $k$=0.12$\epsilon /$\AA $^2$,
i.e. of order 0.4 N/m,  which is typical for atomic force microscopy. 
The outer end of one spring is held constant whereas the outer end
of the other is pulled along the initial end-to-end vector.
Our results are shown for the pulling speed of 0.005 \AA $/\tau$
which corresponds to $7 \times 10 ^6$ nm/s. Even though this speed 
is  3 orders of magnitude faster than in experiments, our 
previous studies \cite{haha,thermtit,homop} indicated only 
small logarithmic
corrections on going to still smaller values of $v_p$.

\section{The force-displacement curves}
 
Stretching of up to five domains of titin 
within the Go model has been analysed in
details in \cite{thermtit}. Here, we present the $F-d$ patterns
for titin to provide a reference for calmodulin.
Figure 3 shows the $F-d$ patterns obtained for one, two, and three
domains of 1tit, linked in tandem, at $\tilde{T}=0$, i.e. when no
thermal fluctuations are taken into account. The multidomain patterns
are essentially a serial repeat of the single domain curve. The single
domain curve has two major force peaks. The first of these has
a height of nearly 4$\epsilon /$\AA and it occurrs due to the
unravelling of the links between the $\beta$-strands that exist
at the opposite 
terminals of the protein. These links are primarily between
the strands A' (amino
acids 11--15), A (amino acids 4--7) and G (amino acids 78--88).
The second major peak is due to breaking the C-F and B-E
links where B, C, E, and F strands correspond to segments 18-25,
32-36, 55-61, and 69-75 respectively. The small hump on the rising
side of the first major peak is due to a rupture 
in the A-B region \cite{APastore}
and it corresponds to the intermediate state that was identified by 
Marszalek et al. \cite{Marszalek0}

Figures 4 and 5 show the
corresponding patterns for $\tilde{T}$ of 0.3 and 0.6. The increase in 
$\tilde{T}$ results in lowering of the force peaks and making them to
occur earlier during the stretching. This is because the thermal
fluctuations provide additional unravelling forces. In the entropic
limit, reached around $\tilde{T}$ of 0.8, there are no identifiable
force peaks and the $F-d$ curves are described by the 
featureless worm-like-chain
model \cite{thermtit,entropic}. In this limit, the domains unravel
simultaneously. At intermediate temperatures, the unravelling is
part serial and part parallel. At $\tilde{T}$=0.3, the stretching of
several domains is predominantly serial in character and the $F-d$ curves
seem to be qualitatively similar to the saw-tooth patterns obtained
experimentally \cite{Gaub,Clarke}. Note that the second major peak force
that has been clearly identified in the $\tilde{T}$=0 trace disappears
at $\tilde{T}=0.3$ except for a weak shadow of it in the first period
of the serial pattern.

The multidomain tandem arrangement is constructed so that
the C-terminal of one domain is connected to the N-terminal of another
by an extra $C^{\alpha} - C^{\alpha}$ bond along the 
end-to-end direction in a single domain.

The $F-d$ curves for calmodulin shown in
Figures 6 through 8 are the analogs of Figures 3 through 5 for titin
and they demonstrate an entirely different behavior.
The single domain curve at $\tilde{T}$=0 displays only minor
force peaks. The biggest force (before the stage of fully stretched
conformation is reached) is only about 1.5$\epsilon /$ \AA
-- less than 40 \% of the maximum force found in titin.
Furthermore, the multidomain curves are not serial repetitions
of the single domain result. Instead, the particular features
in the plot get effectively multiplied in segments, indicating a
large degree of parallelism in the unwinding. Our studies \cite{oldo}
of two helices connected in series have indicated that they unwind
in parallel. Thus the behavior found for calmodulin echos this
finding.

An increase in the temperature causes effects which are consistent
with the general scenario \cite{entropic} -- the peaks get lower and
become less resolved. The interesting part is that, for calmodulin,
the peaks at $\tilde{T}$=0.3 are so inconspicuous that the curves look
as though the system was almost in the entropic/worm-like-chain limit.
This limit is fully achieved at $\tilde{T}$=0.6, as demonstrated
in Figure 8.
The nearly featureless nature of the experimental $F-d$ curves taken
at the room temperature \cite{Progress} 
is consistent with the $\tilde{T}=0.3$
finding based on the Go model.

\section{Scenarios of unfolding for calmodulin}

Insights into unfolding can be obtained by looking
at the snapshots of the process. Figure 9 shows the snapshots for
1cfc at $\tilde{T}$=0. Similar to titin, the system starts unwinding
by breaking bonds that connect the terminal segments. 
However, the breakage involves a much smaller force. The
next stage involves
separation of helices which are parallel to each other, and finally
unwinding of the helices themselves. Figure 10 shows that
the stretching of two domains engages both of them from the
early stages on.

A convenient way to describe the unfolding process is by providing
the 'scenario diagrams' in which distances, $d_u$ at which specific
contacts break are plotted against the contact order, i.e. against 
the sequential distance $s = |j-i|$ between the contact making
amino acids $i$ and $j$. At $\tilde{T}$=0, there is a unique distance
at which a contact breaks. At finite temperatures, contacts may reform
so a meaningful definition of $d_u$ is through a distance
at which the contact exists for the last time. The technical criterion
for the existence of a contact is that the distance between the
amino acids involved is less than 1.5 $\sigma _{ij}$ 
\cite{Hoang,Hoang1,thermtit}.

Figures 11 and 12 show the stretching scenarios for 1cfc at $\tilde{T}=0$
and 0.3 respectively. Six sets of interactions are represented by 
polygonal or circular symbols as indicated in the figures whereas the
remaining interactions are marked by the crosses. For instance, the
interactions between the 6--17 and 65--75 amino acids in two
$\alpha$-helices are shown as the open circles and marked as
H(6-17)--H(65-75). The other character symbols follows the conventions
of the Protein Data Bank: E is an extended $\beta$-strand and S is a bend.
The two scenarios are rather similar to each. For instance, in both
cases stretching is initiated in the terminal region indicated
by CN at the high end of the contact order.
However, 
the ordering of certain events is switched in time. For instance, the
rupturing of the H(82-92)--H(102-111) contacts (the triangles)
at $\tilde{T}$=0 takes place later than that of the H(6-17)--H(65-75)
(the open circles) whereas at $\tilde{T}=0.3$ the opposite 
ordering takes place. Furthermore, the
$\tilde{T}=0.3$ events are more noticeably accumulated
into well defined stripes
compared to the bigger scatter and single event resolution seen in 
the $\tilde{T}$=0 plot.

We now focus on the room-temperature-like case of $\tilde{T}$=0.3
and consider the stretching scenarios for several domains of calmodulin.
Figure 13 refers to the case of two domains. The star symbols denote
the contacts in the first domain whereas the squares to the second domain.
There is a clear intermixing of the symbols which indicates
a non-serial character of unfolding. In particular, the contacts
in the near-terminal (CN) regions unwind nearly simultaneously in 
both domains - in a marked contrast to what happens in two titin
domains \cite{thermtit}. An even heavier mixing takes place in the
case of three domains, as shown in Figure 14.

In summary, the Go-like models can capture the experimentally found
difference in the elastic behavior between calmodulin and titin
and can elucidate the microscopic picture of the events in stretching.
The force-displacement patterns are sensitive to temperature
and acquire the worm-like-chain behavior in the entropic limit.
Unwinding of modular proteins does not need be serial in nature
and calmodulin provides a clear example of such a non-seriality.
Its source is the mostly $\alpha$ character of the protein.

The Author thanks T. X. Hoang and M. O. Robbins for discussions and
collaboration. He also thanks P. E. Marsza{\l}ek for his comments
on the manuscript.
This work was funded by Polish Ministry of Science, project 2 P03B 032 25,
and the European program IP NAPA through Warsaw University of Technology.

\newpage
%
%

\begin{figure}
\epsfxsize=6in
\centerline{\epsffile{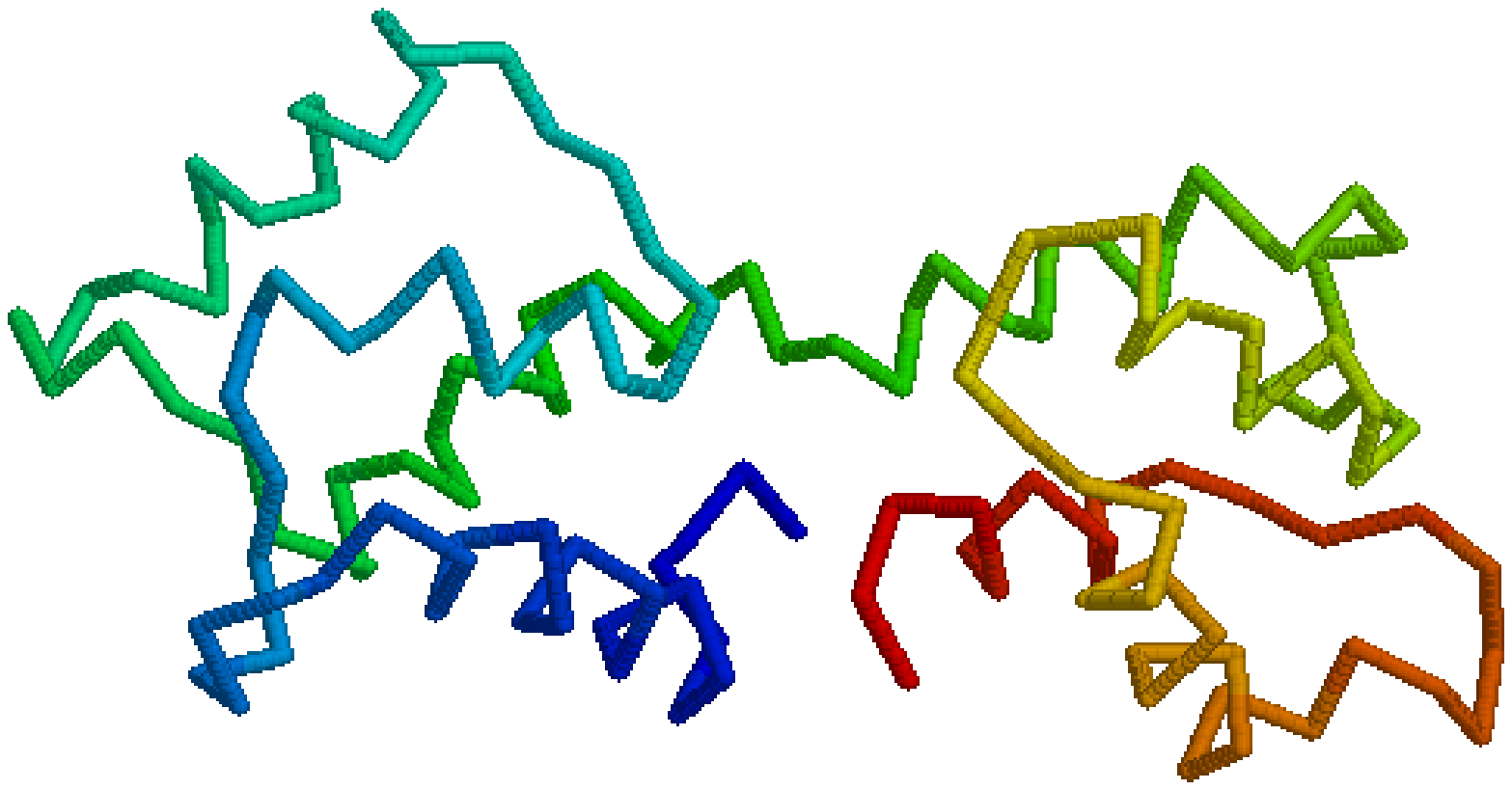}}
\caption{The backbone representation of the 1cfc structure of calmodulin.}
\end{figure}

\begin{figure}
\epsfxsize=6in
\centerline{\epsffile{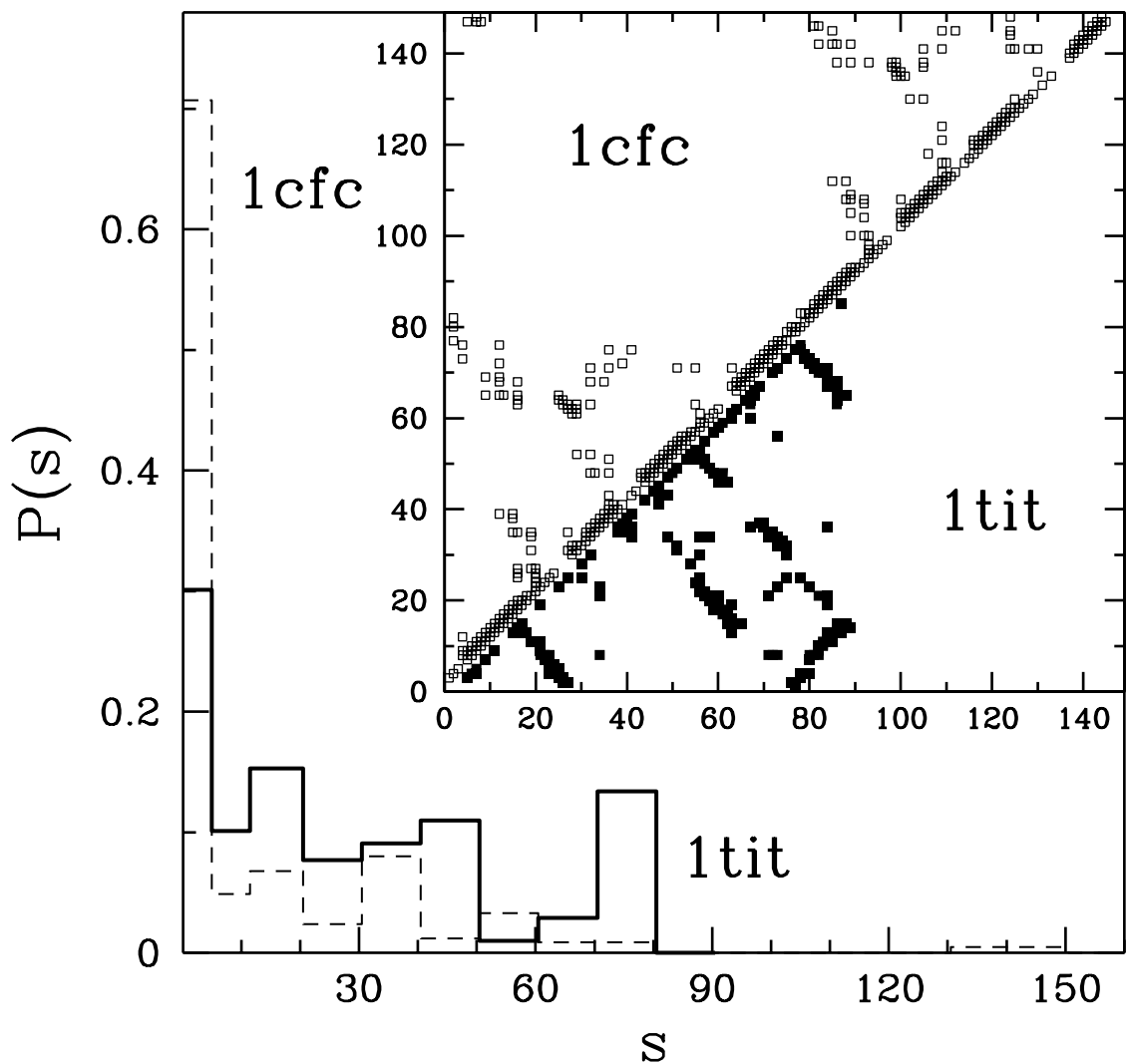}}
\caption{The distribution of sequential distances in the native contacts.
The solid line is for 1tit and the broken line is 1cfc. 
The bin sizes generally correspond to the distance of 10 except for the
very first bin which counts contacts of length smaller or equal to 4
and the second bin which counts contacts with the distance
bigger than 4 but not exceeding 10.
The inset shows
the corresponding contact maps. The contact maps are symmetric and only
a half is shown for each protein.}
\end{figure}

\begin{figure}
\epsfxsize=6in
\centerline{\epsffile{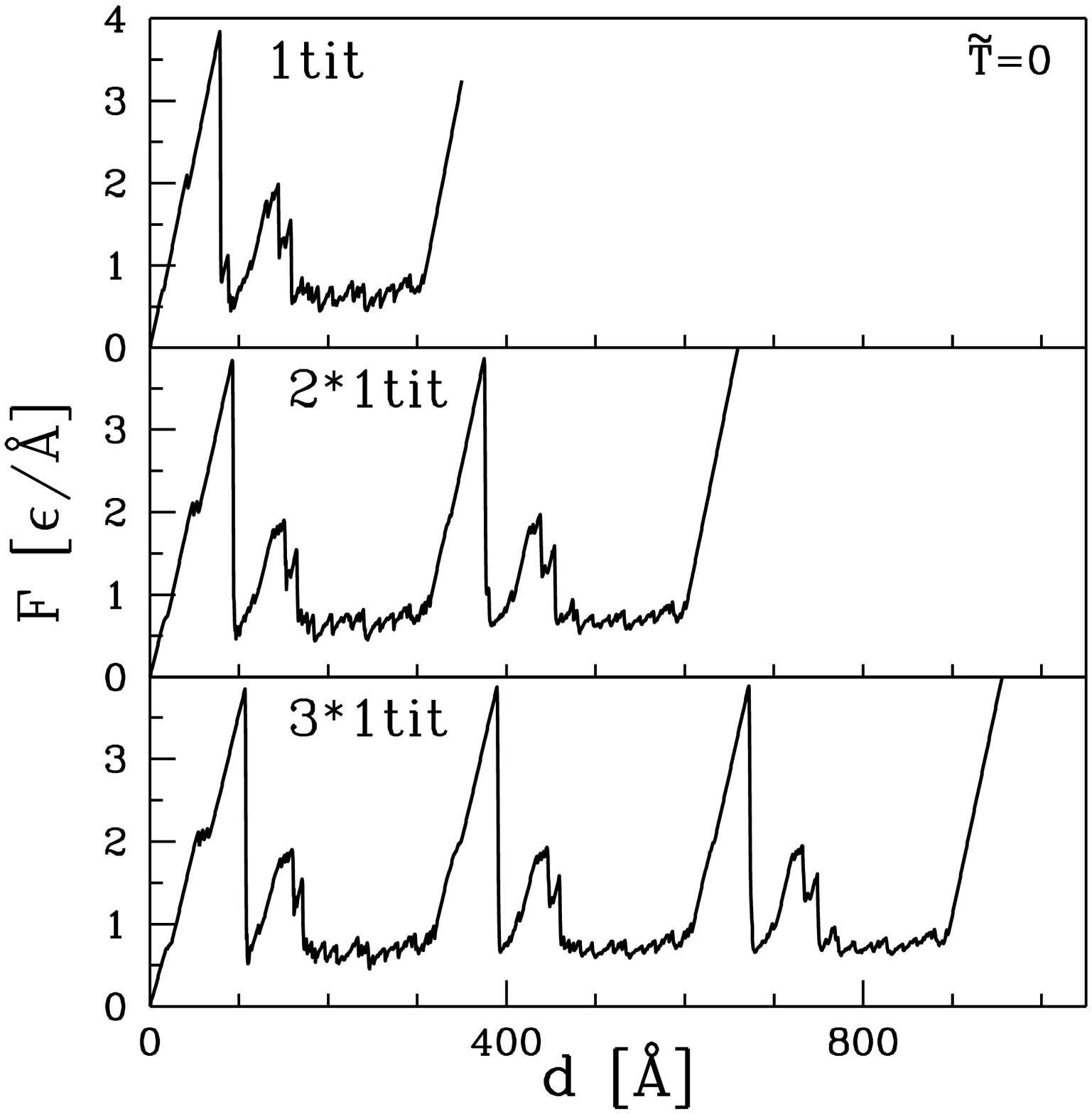}}
\vspace*{3cm}
\caption{The force-displacement curves for one, two, and three
domains of 1tit, top to bottom respectively, for $\tilde{T}$=0.}
\end{figure}

\begin{figure}
\epsfxsize=6in
\centerline{\epsffile{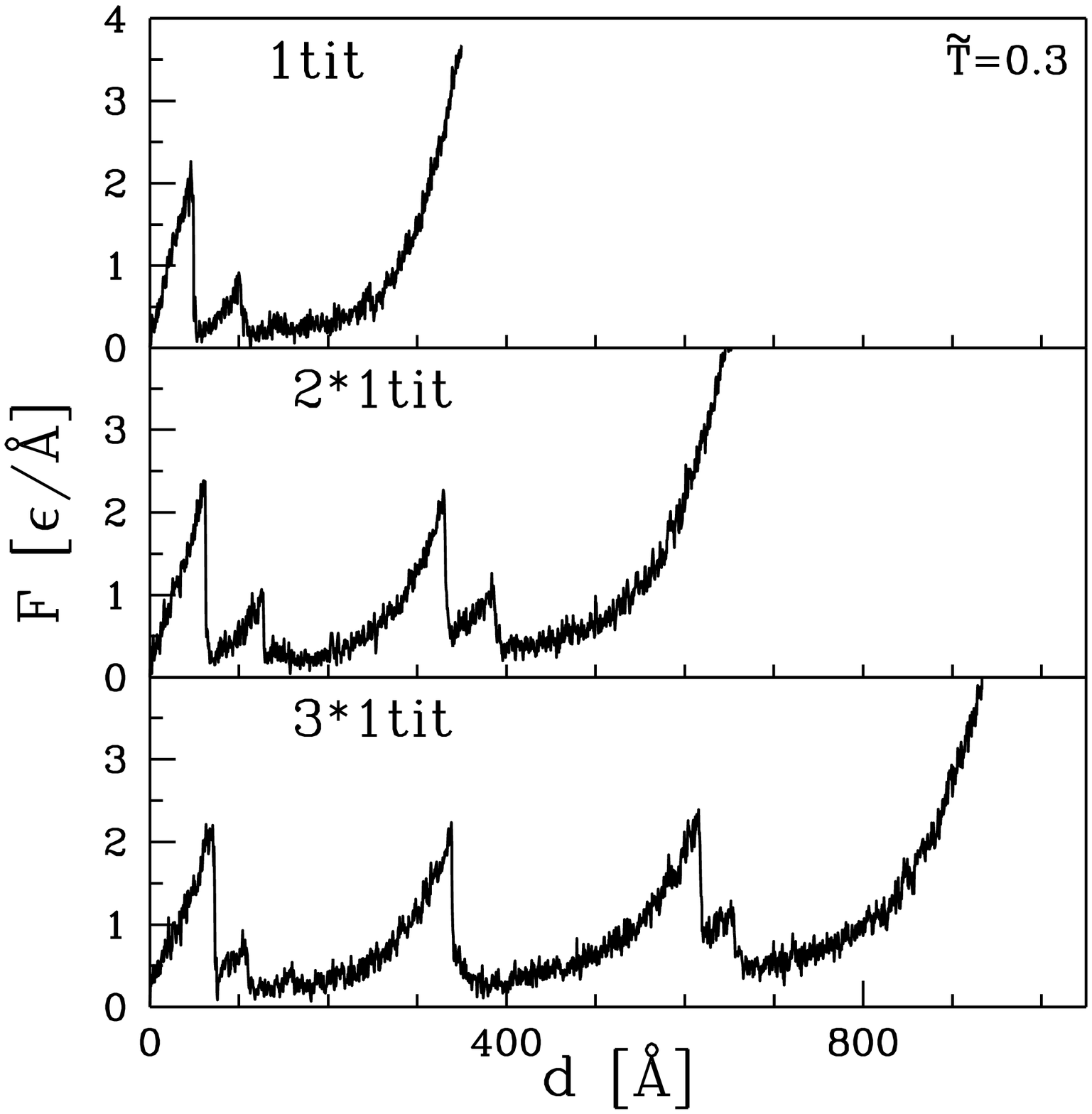}}
\vspace*{3cm}
\caption{Same as in Figure 3 but for $\tilde{T}$=0.3.}
\end{figure}

\begin{figure}
\epsfxsize=6in
\centerline{\epsffile{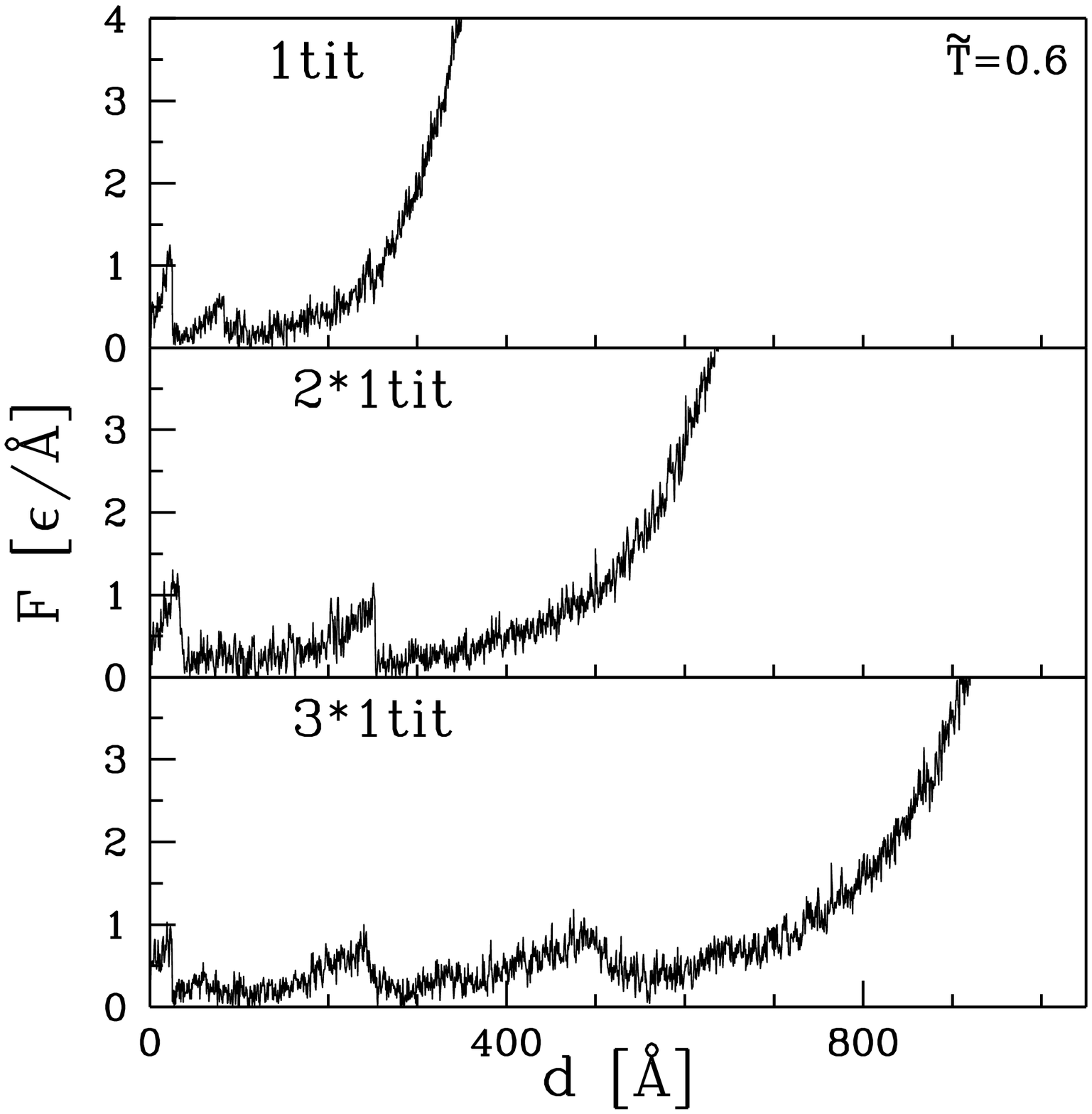}}
\vspace*{3cm}
\caption{Same as in Figure 3 but for $\tilde{T}$=0.6.} 
\end{figure}

\begin{figure}
\epsfxsize=6in
\centerline{\epsffile{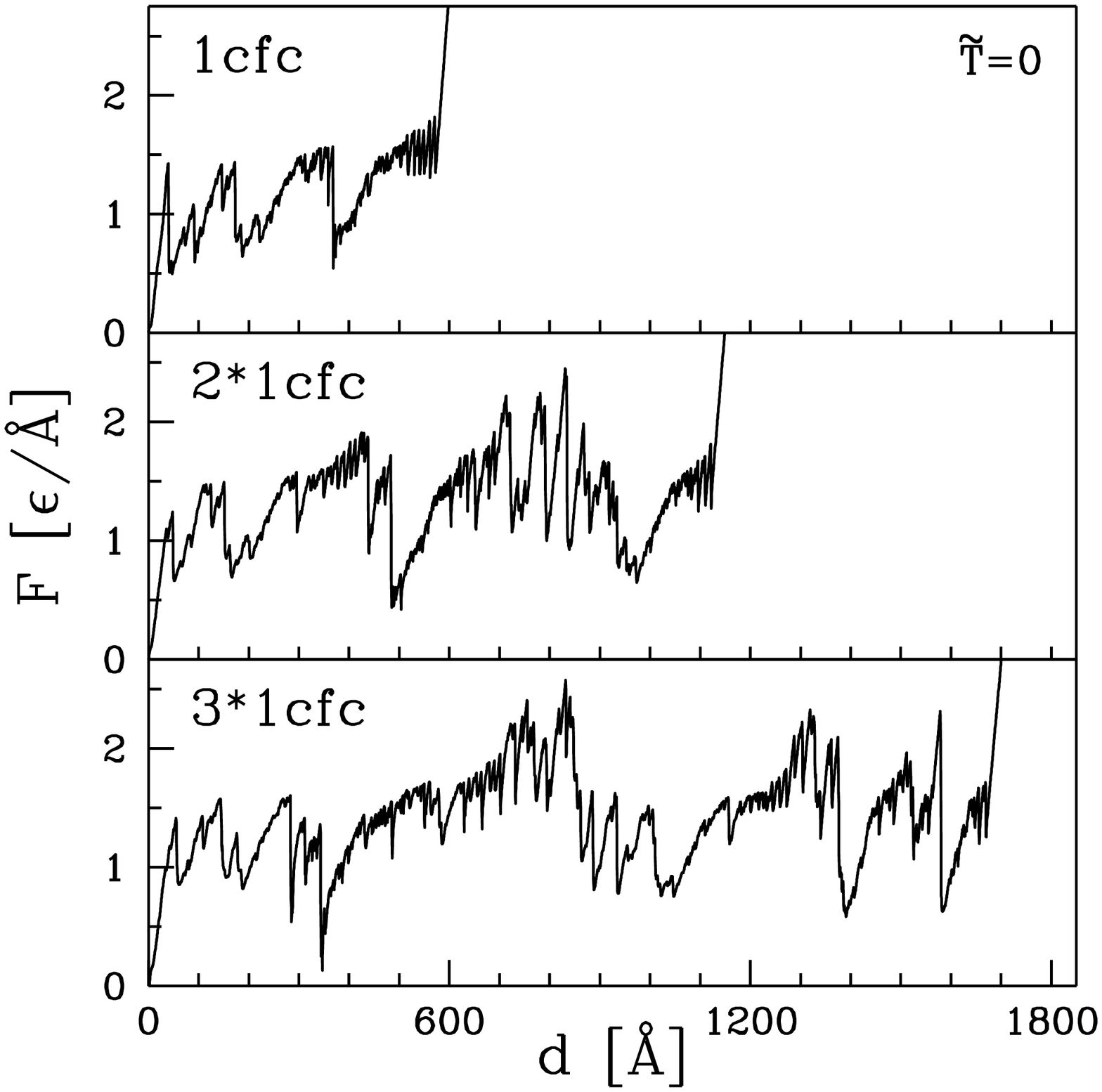}}
\vspace*{3cm}
\caption{The force-displacement curves for one, two, and three domains 
of 1cfc, top to bottom respectively, for $\tilde{T}$=0.}
\end{figure}

\begin{figure}
\epsfxsize=6in
\centerline{\epsffile{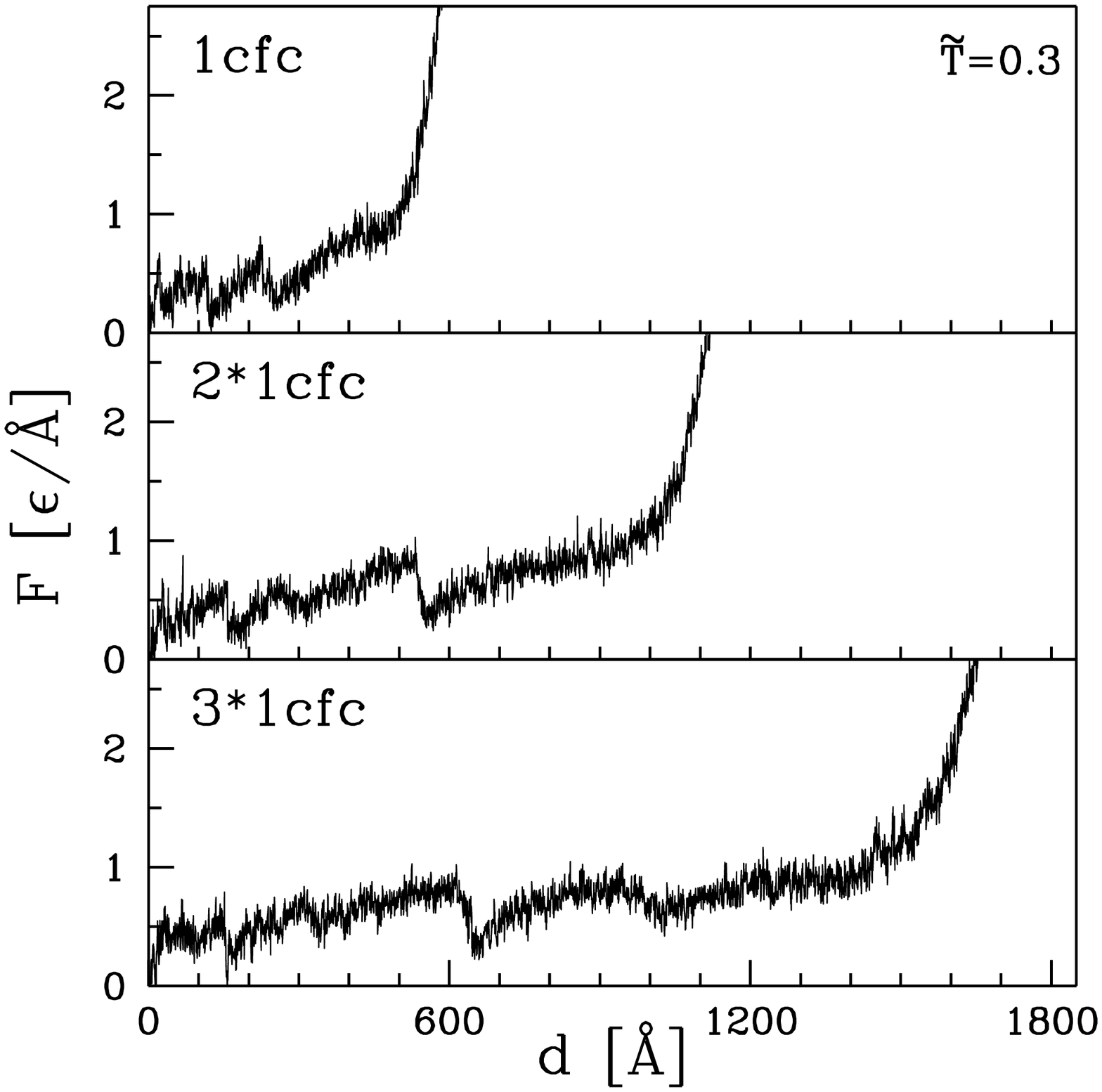}}
\vspace*{3cm}
\caption{Same as in Figure 6 but for $\tilde{T}=0.3$.}
\end{figure}

\begin{figure}
\epsfxsize=6in
\centerline{\epsffile{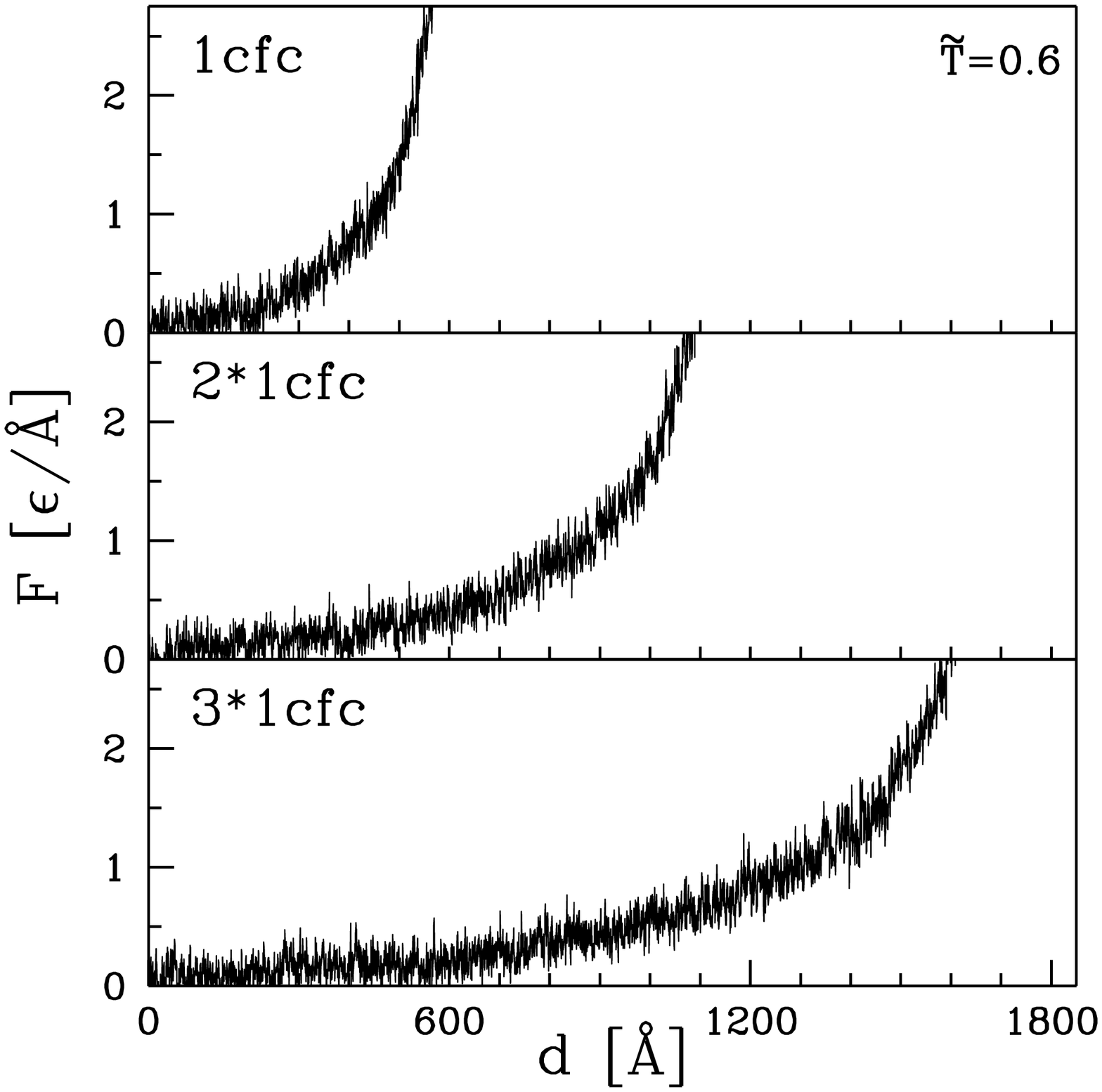}}
\vspace*{3cm}
\caption{Same as in Figure 6 but for $\tilde{T}=0.6$.}
\end{figure}

\begin{figure}
\epsfxsize=6in
\centerline{\epsffile{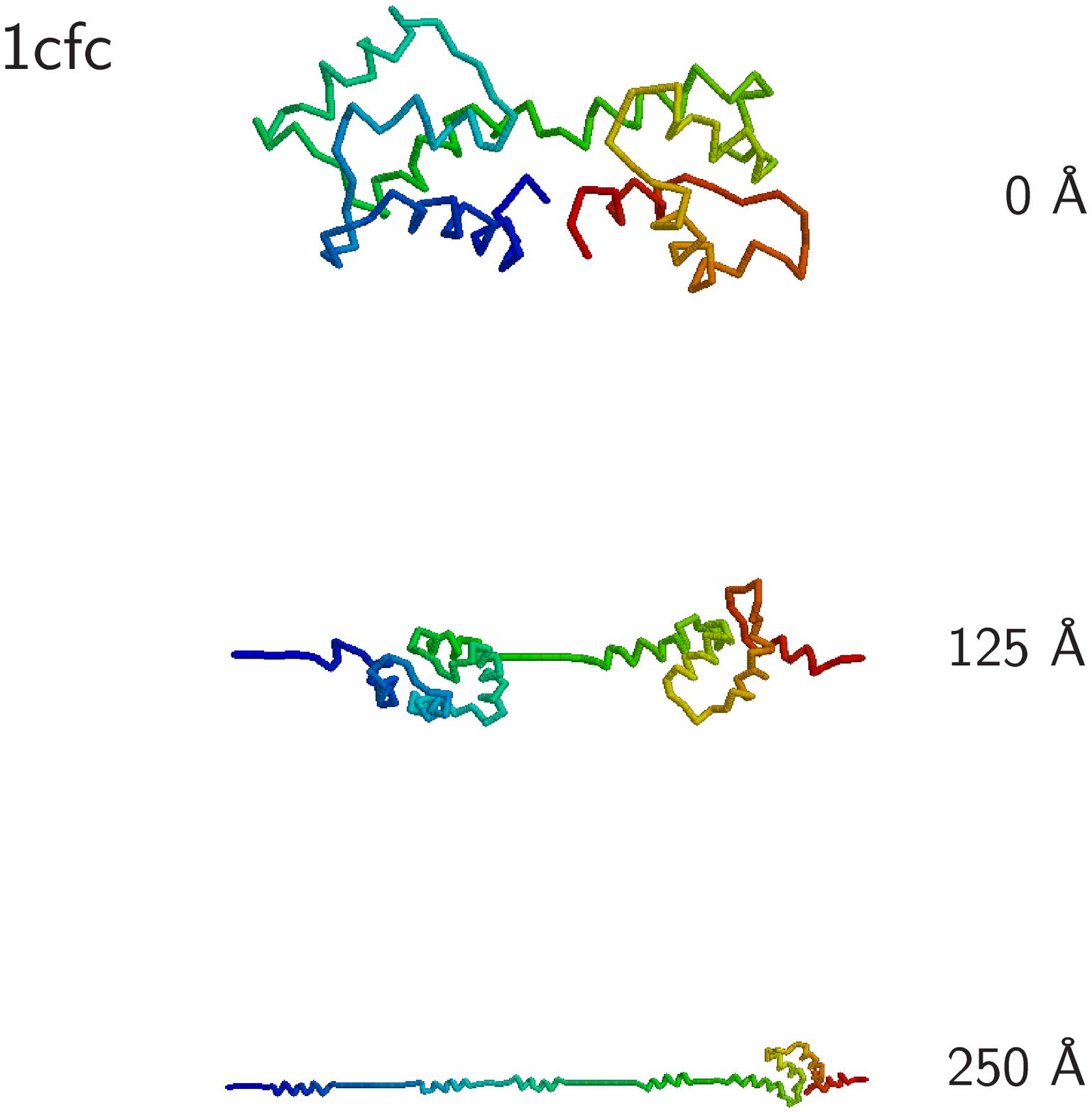}}
\vspace*{3cm}
\caption{The conformations of 1cfc during unfolding at $\tilde{T}$=0.
The labels on the right hand side indicate the corresponding
values of $d$.}
\end{figure}

\begin{figure}
\epsfxsize=6in
\centerline{\epsffile{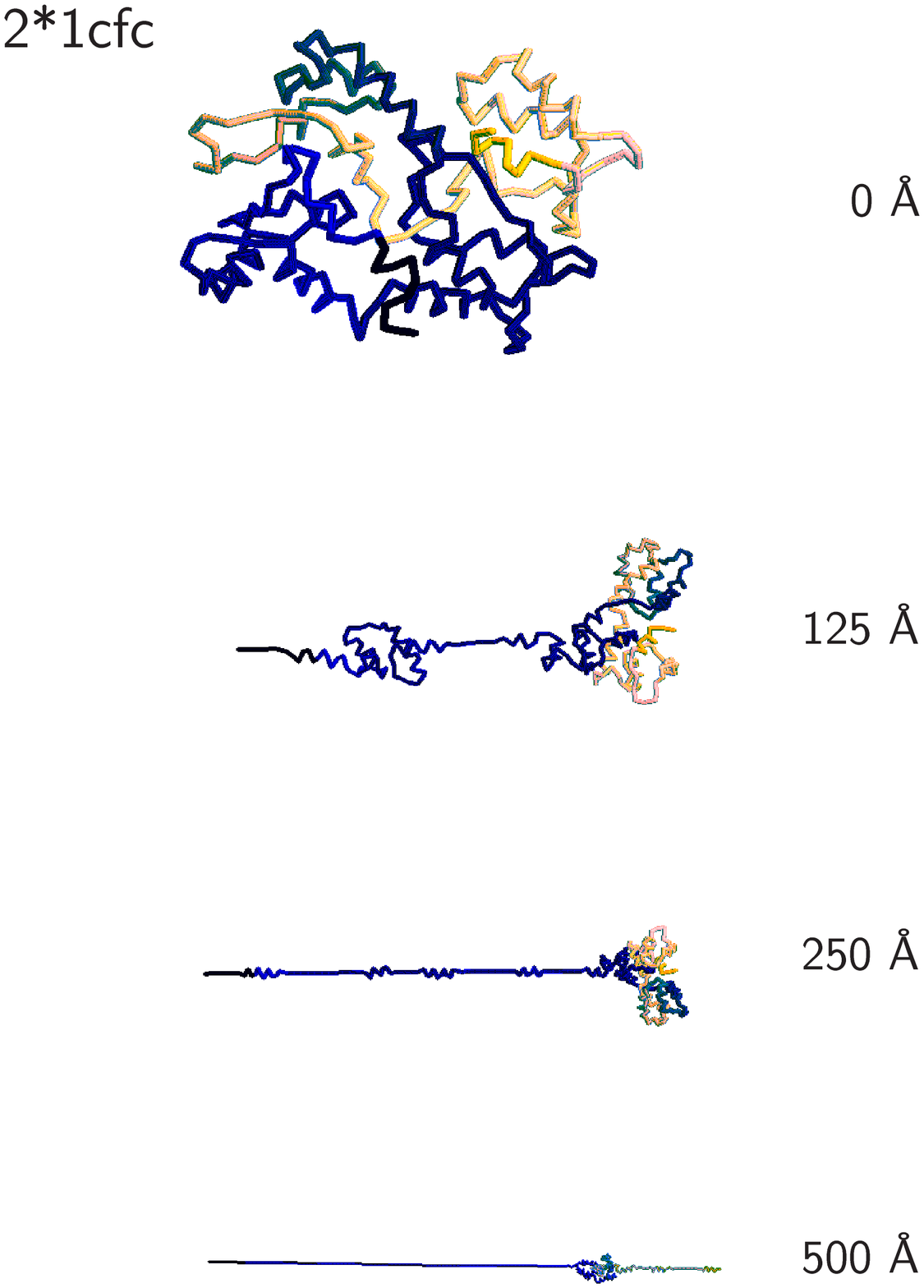}}
\vspace*{3cm}
\caption{Same as in Figure 9 but for two 1cfc domains connected in
tandem.}
\end{figure}

\begin{figure}
\epsfxsize=6in
\centerline{\epsffile{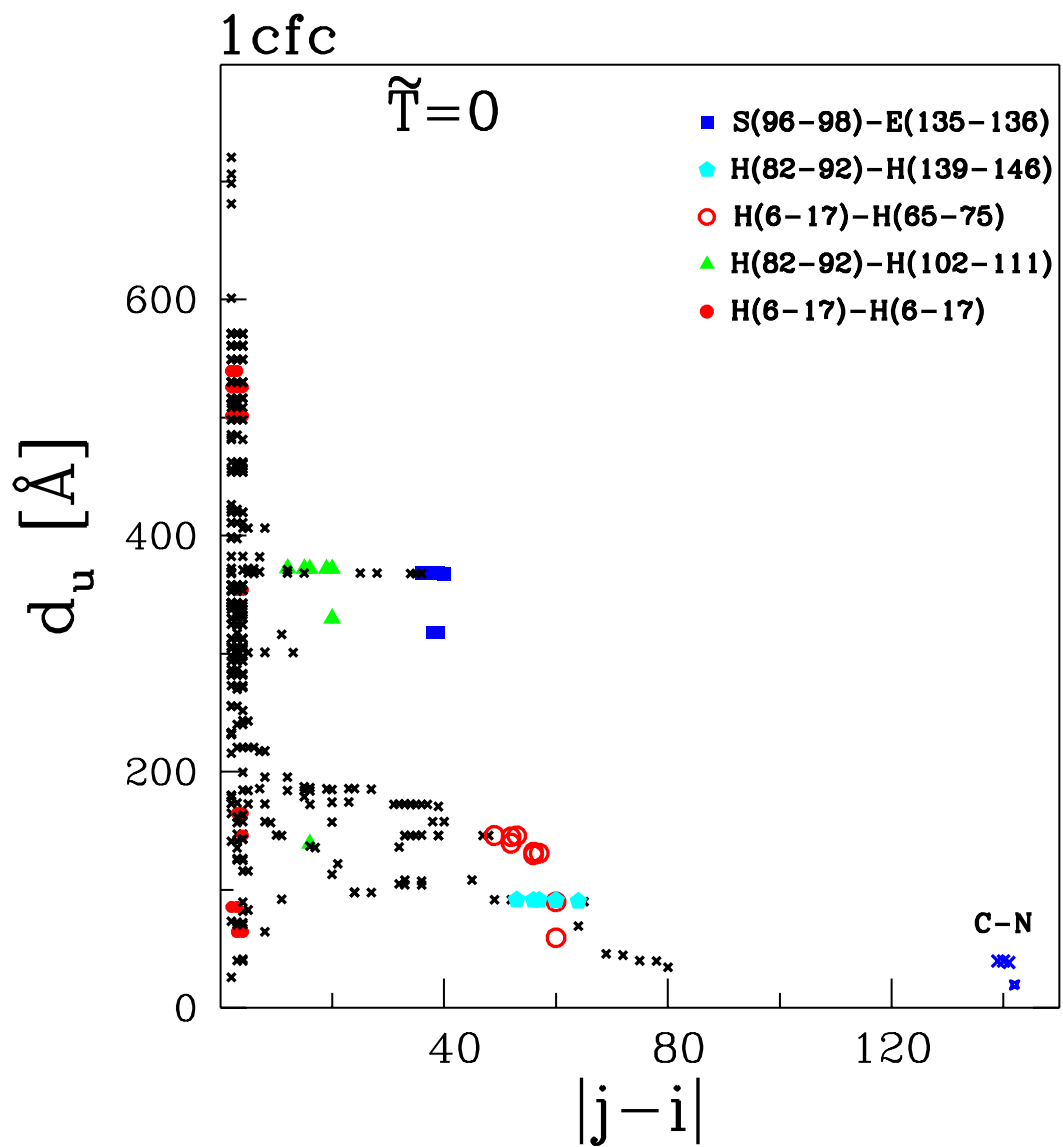}}
\vspace*{3cm}
\caption{The scenario diagram for stretching of 1cfc at 
$\tilde{T}$=0. The symbols are explained in the main text.}
\end{figure}

\begin{figure}
\epsfxsize=6in
\centerline{\epsffile{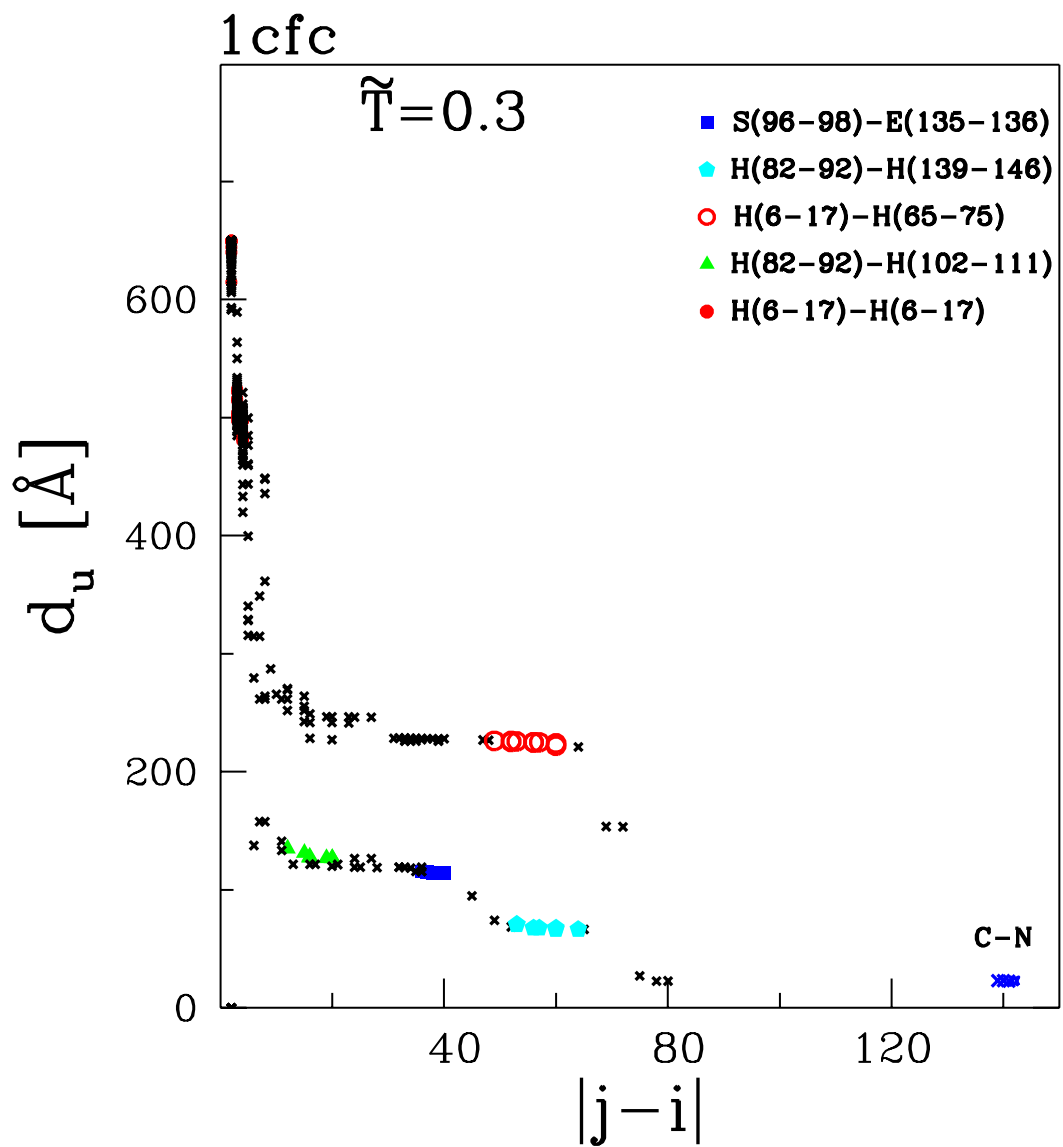}}
\vspace*{3cm}
\caption{Same as in Figure 11 but for $\tilde{T}=0.3$.
The data points are for a single trajectory.}
\end{figure}

\begin{figure}
\epsfxsize=6in
\centerline{\epsffile{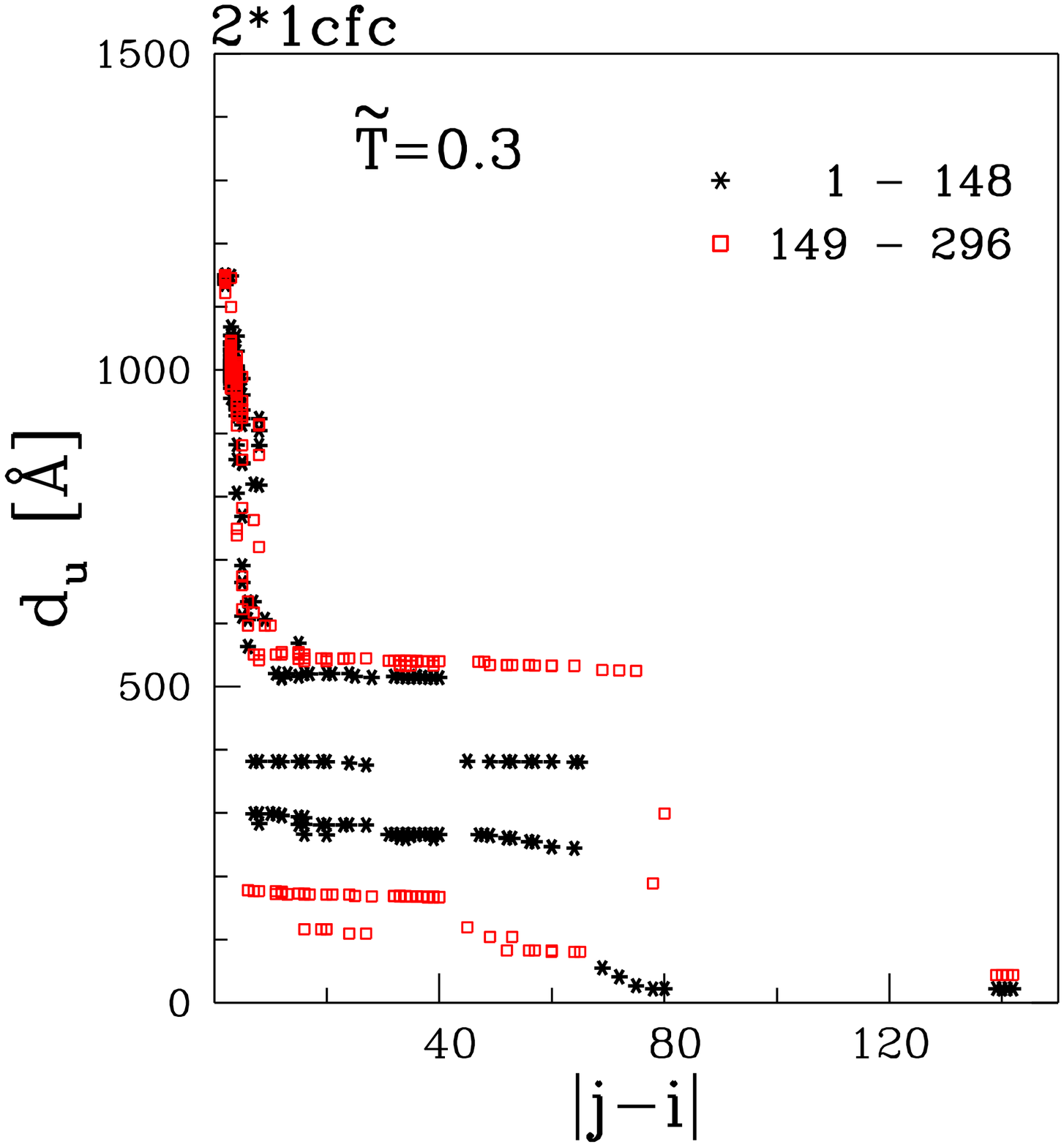}}
\vspace*{3cm}
\caption{The scenario diagram for two domains of 1cfc at 
$\tilde{T}=0.3$.}
\end{figure}

\begin{figure}
\epsfxsize=6in
\centerline{\epsffile{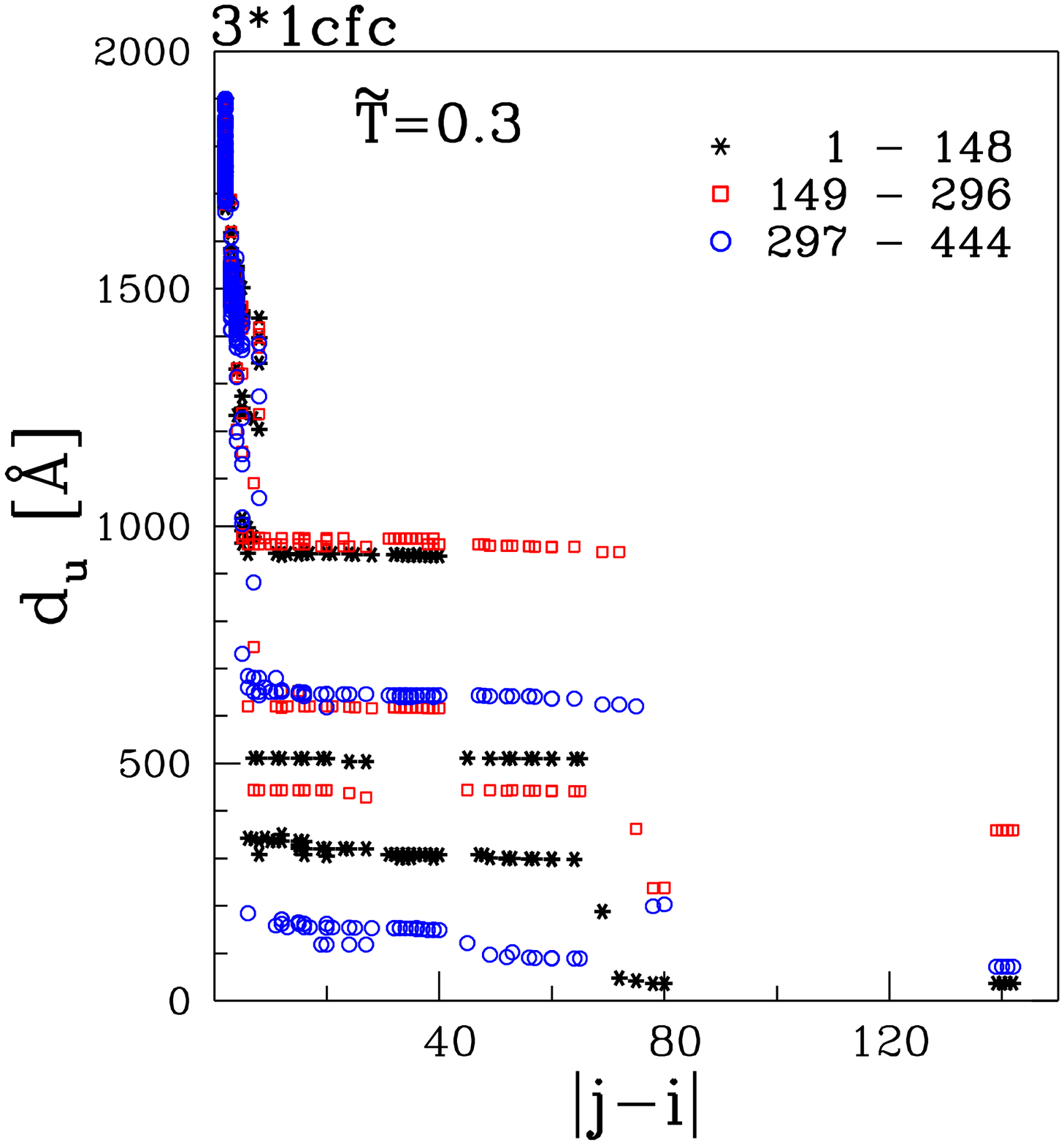}}
\vspace*{3cm}
\caption{The scenario diagram for three domains of 1cfc at 
$\tilde{T}=0.3$.}
\end{figure}


\begin{thebibliography}{00}




\bibitem{Florin}
E. L. Florin, V. T. Moy, and H. E. Gaub,
Science {\bf 264}, 415-417 (1994)

\bibitem{Grubmueller}
H. Grubmuller, B. Heymann, and P. Tavan,
Science {\bf 271}, 997-999 (1996).

\bibitem{Smith}
S. B. Smith, Y. Cui, and C. Bustamante, 
Science {\bf 271}, 795-799 (1996).

\bibitem{Cluzel}
A. Cluzel, A. Lebrun, C. Heller, R. Lavery, J. L. Viovy, 
D. Chatenay, and F. Caron,
Science {\bf 271}, 792-794 (1996).

\bibitem{Heslot}
U. Bockelmann, B. Essevaz-Roulet, and F. Heslot,
Phys. Rev. Lett. {\bf 79}, 4489-4492 (1997).

\bibitem{Gaub}
M. Rief, M. Gautel, F. Oesterhelt, J. M. Fernandez, and H. E. Gaub,
Science {\bf 276}, 1109-1112 (1997).

\bibitem{Simmons}
L. Tskhovrebova, K. Trinick, J. A. Sleep, and M. Simmons,
Nature {\bf 387} 308-312 (1997).

\bibitem{Kellermayer}
M. S. Z. Kellermayer, S. B. Smith, H. L. Granzier, and C. Bustamante, 
Science {\bf 276} 1112-1116 (1997).

\bibitem{Clarke}
S. B. Fowler, R. B. Best, J. L. Toca Herrera, T. J. Rutherford, A. Steward,
E. Paci, M. Karplus, and J. Clarke,
J. Mol. Biol. {\bf 322} 841-849 (2002).

\bibitem{Marszalek0}
P. E. Marszalek, H. Lu, H. B. Li, M. Carrion-Vazquez, A. F. Oberhauser, 
K. Schulten, and J. M. Fernandez,
Nature {\bf 402} 100-103 (1999).

\bibitem{Marszalek}
M. Carrion-Vasquez, A. F. Oberhauser, S. B. Fowler, P. E. Marszalek,
S. E. Broedel, J. Clarke, and J. M. Fernandez,
Proc. Natl. Acad. Sci. USA {\bf 96} 3694-3699 (1999).

\bibitem{Guan}
Z. B. Guan, J. T. Roland, J. Z. Bai, S. X. Ma, T. M. McIntire, 
and M. Nguyen,
J. Am. Chem. Soc. {\bf 126}, 2058-2065 (2004).

\bibitem{Progress} M. Carrion-Vazquez, A. F. Oberhauser, T. E. Fisher,
P. E. Marszalek, H. LI, and J. M. Fernandez,
Prog. Biophys. Mol. Biol. {\bf 74} 63-91 (2000).

\bibitem{Carrion-Vazquez}
M. Carrion-Vazquez, H. Li, H. Lu, P. E. Marszalek, A. F. Oberhauser,
and J. M. Fernandez,
Nat. Struct. Biol. {\bf 10} 738-743 (2003).

\bibitem{Yang}
G. Yang, C. Cecconi, W. A. Baase, I. R. Vetter, W. A. Breyer, J. A. Haack,
B. W. Matthews, F. W. Dahlquist, and C. Bustamante,
Proc. Nat. Acad. Sci. {\bf 97}, 139-144 (2000).

\bibitem{entropic}
M. Cieplak, T. X. Hoang, and M. O. Robbins,
Phys. Rev. E. {\bf 69}, 011912 (2004).

\bibitem{Janoviak}
H. Janovjak, M. Kessler, D. Oesterhelt, H. Gaub, and D. J. Mueller,
The EMBO Journal {\bf 22}, 5220-5229 (2003).

\bibitem{thermtit}
M. Cieplak, T. X. Hoang, and M. O. Robbins,
Thermal effects in stretching of Go-like models of titin and
secondary structures,
Proteins: Struct. Funct. Bio. {\bf 56}, 285-297 (2004).

\bibitem{PDB}
F. C. Bernstein, T. F. Koetzle, G. J. B. Williams,
E. F. Meyer Jr., M. D. Brice, J. R. Rodgers, O. Kennard, T. Shimanouchi,
and M. Tasumi,
J. Mol. Biol. {\bf 112}, 535-542 (1977).

\bibitem{Pastore}
Improta S, Politou AS, Pastore A.
Immunoglobulin-like modules from titin I-band: extensible components
of muscle elasticity.
Structure 1996;15:323-327.

\bibitem{Lu}
Lu H, Schulten K.
Steered molecular dynamics simulation of conformational changes of
immunoglobulin domain I27 interprete atomic force microscopy observations
Chemical Physics 1999;247:141-153.

\bibitem{geom} See, e.g., C. Micheletti, J. R. Banavar, A. Maritan,
and F. Seno,
Phys. Rev. Lett. {\bf 82} 3372-3375 (1999).

\bibitem{tubes} A. Maritan, C. Micheletti, A. Trovato, and J. R. Banavar,
Nature {\bf 406} 287 (2000).

\bibitem{tubes1} J. R. Banavar and A. Maritan,
Rev. Mod. Phys. {\bf 75} 23-34 (2003).

\bibitem{Du}
R. Du, V. S. Pande, A. Y. Grosberg, T. Tanaka, and E. I. Shakhnovich,
J. Chem. Phys. {\bf 111} 10375-10380 (1999).

\bibitem{Unger}
R. Unger and J. Moult,
J. Mol. Biol. {\bf 259} 988-994 (1996).

\bibitem{Plaxco}
K. W. Plaxco, K. T. Simons, and D. Baker,
J. Mol. Biol. {\bf 277} 985-994 (1998).

\bibitem{Plaxco1}
K. W. Plaxco, K. T. Simons, I. Ruczinski, D. Baker,
Biochemistry {\bf 39} 11177-11183 (2000).

\bibitem{Goabe}
H. Abe, N. Go,
Biopolymers {\bf 20}, 1013-1031 (1981).

\bibitem{Stakada}
S. Takada,
Proc. Natl. Acad. Sci. USA {\bf 96}, 11698-11700 (1999).

\bibitem{Hoang}
T. X. Hoang and M. Cieplak,
J. Chem. Phys. {\bf 112}, 6851-6862 (2000).

\bibitem{Hoang1}
T. X. Hoang and M. Cieplak,
J. Chem. Phys. {\bf 113}, 8319-8328 (2001).

\bibitem{biophysical}
M. Cieplak and T. X. Hoang,
Biophys. J. {\bf 84} 475-488 (2003).

\bibitem{Tsai}
J. Tsai, R. Taylor, C. Chothia, and M. Gerstein, J. Mol. Biol.
{\bf 290}, 253-266 (1999).

\bibitem{prion}
G. Settanni, T. X. Hoang, C. Micheletti, and A. Maritan,
Biophys. J. {\bf 83}, 3533-3541 (2002).

\bibitem{Veitshans}
T. Veitshans, D. Klimov, and D. Thirumalai,
Folding Des. {\bf 2}, 1-22 (1997).

\bibitem{haha}
M. Cieplak, T. X. Hoang, and M. O. Robbins,
Proteins: Function, Structure, and Genetics {\bf 49}, 114 - 124 (2002).

\bibitem{homop}
M. Cieplak, T. X. Hoang, and M. O. Robbins,
Phys. Rev. E {\bf 70} 011917 (2004).

\bibitem{APastore}
M. Cieplak, A. Pastore, and T. X. Hoang,
J. Chem. Phys. (in press).

\bibitem{oldo}
M. Cieplak, T. X. Hoang, and M. O. Robbins,
Proteins: Function, Structure,
and Genetics {\bf 49}, 104 - 113 (2002).



\end{thebibliography}
\end{document}